\documentstyle[12pt]{article}

\renewcommand{\thefootnote}{\fnsymbol{footnote}}
\newcommand{\r}[1]{(\ref{#1})}

\begin{document}
\thispagestyle{empty}
\newlength{\defaultparindent}
\setlength{\defaultparindent}{\parindent}

\begin{center}

{\large{
Maxwell--Chern-Simons gauged non-relativistic O(3) model with self-dual
vortices }}

\vspace{1.5cm}

{\large D. H. Tchrakian} \footnote{Permanent address: Department of
Mathematical Physics, St Patrick's College, Maynooth, Ireland. Supported in
part by the TMR grant ERBFMBICT960905} and {\large T.N. Tomaras}

{\small Department of Physics, and Institute of Plasma Physics, University
of Crete\\
and Research Center of Crete\\
P.O.Box {\small 2208}, {\small 710 03} Heraklion, Greece \\}

\end{center}
\bigskip
\bigskip
\bigskip

\begin{abstract}
A non-relativistic version of the 2+1 dimensional gauged Chern-Simons
$O(3)$ sigma model, augmented by a Maxwell term, is presented and shown
to support topologically stable {\it static} self-dual vortices. Exactly like
their counterparts of the ungauged model, these vortices are shown to
exhibit Hall behaviour in their {\it dynamics}. \end{abstract}

\vfill
\setcounter{page}0
\renewcommand{\thefootnote}{\arabic{footnote}} \setcounter{footnote}0
\newpage

\newcommand{\ra}{\rightarrow}

\newcommand{\dd}{\mbox{d}}
\newcommand{\ee}{\end{equation}}
\newcommand{\be}{\begin{equation}}
\newcommand{\ii}{\mbox{ii}}
\newcommand{\pa}{\partial}
\newcommand{\vep}{\varepsilon}
\newcommand{\bfR}{{\bf R}}
\newcommand{\lm}{\lambda}

\pagestyle{plain}

\parskip0.5truecm

\section{Introduction}

Non-relativistic field theories in 2+1 dimensions supporting vortex
solutions are important in planar physics. Vortices feature in the $O(3)$
ferromagnet Landau-Lifshitz model \cite{PT}, the
Ginzburg-Landau model of superconductivity \cite{PTST}, and in the
charged
quantum fluid \cite{ST}.
The last two theories are $U(1)$ gauged theories while the former is
ungauged. It is
therefore interesting to consider the $U(1)$ gauged version of the
non-relativistic $O(3) \: \: \sigma$ model,
the Landau-Lifshitz model of ferromagnetism,
which is presented below.

The $U(1)$ gauging we perfom employs both Maxwell and Chern-Simons dynamics.
The inclusion of the Chern-Simons term enables the description of anyonic
dynamics,
but in addition to this, its presence leads to two very interesting features of
the model. The first of these is that in the
limit of vanishing gauge coupling constant the resulting $U(1)$ gauged
$O(3)\: \: \sigma$
model reduces to the Landau-Lifshits model of ferromagnetic materials \cite{PT}.
The second of these is that the resulting
system supports self-dual vortex solutions, which is a very useful
mathematical feature.
Concerning the inclusion of the Maxwell term, this turns out to be
necessary for the
topological stability of the vortices.

Non-relativistic Chern-Simons vortices in 2+1 dimensions were first introduced by Jackiw
and Pi \cite{JP} in the context of a $U(1)$ gauged non-linear Schr\"odinger
equation, which apart from the self-interaction potential of the matter
field resembles the Ginsburg-Landau equation. The theory supported self-dual
solutions in the static limit, which were characterised by the large
distance asymptotic property $\lim_{r \rightarrow \infty} |\Psi| =0$
of the complex valued matter field $\Psi$,
according to which these vortices do not have a non-trivial vacuum condensate.
One consequence of this is, that
the energy of their solutions is not bounded from below by a topological charge
and hence is not guaranteed to be stable. To change
this situation, namely to allow for vortices exhibiting a non-trivial
condensate, Barashenkov and Harin \cite{BH} modified the model of Refs.
\cite{JP}. One of the more remarkable features of the modified model of
Refs. \cite{BH}, which features {\it both} Maxwell {\it and} Chern-Simons
dynamics, is that the static vortices are topologically stable and self-dual,
and the resulting model agrees with the non-relativistic Ginzburg-Landau theory,
with the Lagrangian augmented by a Chern-Simons term. The self-dual vortex
solutions
of Ref. \cite{BH} were
previously found by Donatis and Iengo \cite{DI}, who worked directly with the
Hamiltonian of the system, where the Chern-Simons term is implicitly
present in the
Lagrangian formulation. Moreover
in Ref. \cite{DI} the chiral aspect of this system was emphasised,
resulting in the
absence of negative (positive) vorticities in the model (cojugate-model). 
We shall
have occasion to discuss this point in our case too. In the present paper, where
we tackle the non-relativistic $U(1)$ gauged $O(3)\: \: \sigma$ model, analogous
to the non-relativistic Ginzburg-Landau model,
we shall follow the Lagrangian formulation of Ref. \cite{BH}.

In a more general context independently of the presence of the Chern-Simons
term,
we discuss the dynamics of these vortices in the framework of the
description given in
Refs. \cite{PT,PTST}. The purpose of this discussion is to highlight the
common features in the non-relativistic dynamics of the vortices of all
these 2+1
dimensional models. Most important amongst these features is the
dependence of the correct definition of the momentum on the definition of
the gauge invariant topological charge density. The prescription
for doing this is a non-trivial matter
and the fact that it turns out to be model-independent is, in our opinion, an
important demonstration of the universality of this prescription given in
Refs. \cite{PT,PTST}.

We present our model, and implicitly give the vortex solutions, below in
Section {\bf 2}.
The dynamics of our vortices is given in Section {\bf 3}, and a brief
summary of our
results is given in Section {\bf 4}.

\section{The model}

In this Section we introduce our model and establish the self-duality equations
minimising the energy.
Since the vortices of our model will turn out to be solutions to the
self-duality
equations to the $U(1)$ gauged $O(3) \;\sigma$ model \cite{S}, 
our vortex solutions
coincide with the latter and hence are implicitly given in this Section.

The model we introduce is described in terms
of the $U(1)$ gauge field $A_{\mu}$ and the scalar field $\phi^a$ subject to
the condition $\phi^a \phi^a =1$, with $\mu=0,i;\: i=1,2$ labeling the
coordinates of the 2+1 dimensional space, and $a=\alpha ,3;\: \alpha =1,2$. A
crucial role will be played by the gauging prescription we employ, which
is the one used by Schroers \cite{S} in the construction of the relativistic
Maxwell--$O(3)$ vortices, and subsequently in the construction of the
relativistic Chern-Simons--$O(3)$ vortices by Ghosh et al \cite{G}, Kimm et
al \cite {KLL} and ourselves \cite{ATY}. The prescription is characterised
by the definition of the covariant derivative $D_{\mu} \phi^a$ as
\begin{equation}
\label{2}
D_\mu \phi^{\alpha} =\partial_\mu \phi^{\alpha} +A_\mu
\varepsilon^{\alpha \beta} \phi^{\beta} ,\qquad D_{\mu} \phi^3
=\partial_{\mu} \phi^3 .
\end{equation}
This prescription of gauging was given employed earlier in Ref. \cite{BT}.
The proposed Lagrangian is
\begin{equation}
\label{3}
\begin{array}{rl}
{\cal L}=& -\frac{\mu^2}{4} F_{\mu \nu} F^{\mu \nu} +\frac{\kappa}{2}
\varepsilon^{\mu \nu \lambda} F_{\mu \nu} A_{\lambda} \\
\\
&+ g(\phi^3)\varepsilon^{\alpha \beta} \phi^{\alpha} D_0 \phi^{\beta}
-\frac{1}{2m} |D_i \phi^a |^2 -U(\phi^3) +\lambda (1- |\phi^a|^2 ) ,
\end{array}
\end{equation}
where $\lambda$ is a Lagrange multiplier.
Both the function $g(\phi^3)$ and the $O(3)$ breaking potential $U(\phi^3)$ will
be fixed by the criteria of topological stability.
The choice
for $g(\phi^3)$ can be made at this stage by requiring its correspondence
with the Landau-Lifshitz theory in the ungauged limit, but we shall instead
derive it below by requiring the existence of static self-dual vortices.

The Gauss Law constraint for \r{3} is
\begin{equation}
\label{4}
\mu^2 \partial_i E_i +\kappa B -g(\phi^3) |\phi^{\alpha}|^2 =0
\end{equation}
where $E_i = - F_{i0}$, and
$B={1\over 2} \varepsilon_{ij} F_{ij}$. The Hamiltonian is
\begin{equation}
\label{5}
{\cal H}=\frac{\mu^2}{2} (E_i^2 + B^2 ) +\frac{1}{2m} |D_i \phi^a |^2 +
U(\phi^3).
\end{equation}
In the static limit \r{4} and \r{5} reduce, respectively, to
\begin{equation}
\label{6}
-\mu^2 \Delta A_0 +\kappa B =g(\phi^3) |\phi^{\alpha}|^2 \end{equation}
\begin{equation}
\label{7}
{\cal H}_{static} =\frac{\mu^2}{2} (\partial_i A_0 )^2 + {\cal H}_0
\end{equation}
with ${\cal H}_0$ given by
\begin{equation}
\label{8}
{\cal H}_0 =\frac{\mu^2}{2} B^2 +\frac{1}{2m} |D_i \phi^a|^2 +U(\phi^3)
\end{equation}

In the temporal gauge $A_0 =0$, the static Hamiltonian \r{7} becomes equal
to the density ${\cal H}_0$ given by \r{8}, and the constraint \r{6} reduces
to
\begin{equation}
\label{8a}
\kappa B = g(\phi^3)|\phi^{\alpha}|^2 =g(\phi^3)(1-(\phi^3)^2).
\end{equation}
The existence of self-dual vortices in the model \r{8}, with a specific choice
of the $O(3)$ breaking potential
\begin{equation}
\label{8b}
U(\phi^3)=U_0 (\phi^3)=\frac{\mu^2}{2} (1-\phi^3)^2 ,
\end{equation}
has been  established in Refs. \cite{S,Y}. The self-duality equations which
minimise
the energy of the Hamiltonian ${\cal H}_0$ given by \r{8} and \r{8b}, are
\begin{equation}
\label{9}
\varepsilon_{ij} D_i \phi^a =\varepsilon^{abc} D_j \phi^b \phi^c
\end{equation}
\begin{equation}
\label{10}
B=\mu (1-\phi^3),
\end{equation}
which for the vortex field configurations in question are satisfied, in
addition to
the Gauss Law constraint \r{8a}. As the number of equations to be
satisfied, \r{8a},
\r{9} and \r{10} exceeds the number of fields $(A_i ,\phi^a)$, it appears
on first
sight that the system is overdetermined. Fortunately however we have not
yet specified
the function $g(\phi^3)$, so we do this such that equations \r{8a} and
\r{10} become
identical, thus reducing the number of the equations to the Bogomol'nyi
equations
\r{9}-\r{10}. This choice is
\[
g(\phi^3)=\kappa \mu (1+\phi^3)^{-1} ,
\]
leading to the final form of the proposed model
\begin{equation}
\label{15}
\begin{array}{rl}
{\cal L} =&-\frac{\mu^2}{4} F_{\mu \nu} F^{\mu \nu} + \frac{\kappa}{2}
\varepsilon^{\mu \nu \lambda} F_{\mu \nu} A_{\lambda} \\
\\
&+\frac{\kappa \mu}{1+\phi^3} \varepsilon^{\alpha \beta} \phi^{\alpha}
D_0 \phi^{\beta}
-\frac{1}{2m} |D_i \phi^a|^2 -\frac{\mu^2}{2} (1-\phi^3)^2 .
\end{array}
\end{equation}
The self-dual vortices supported by the static Hamiltonian \r{8} and \r{8b},
are well known from the work of Ref. \cite{S}, where it was shown that vortex
configurations of vorticities $N\ge 2$ satisfy the Bogomol'nyi equations
\r{9}-\r{10}.

It is in order at this point to remark that we can acheive the topological
lower bound
by saturating the {\it self-duality} equations \r{9}-\r{10}, but unlike in
the case of
the relativistic $U(1)$ gauged $O(3)$ sigma models \cite{S,G,KLL,ATY},
it is not possible to
acheive this by imposing instead the {\it anti--self-duality} equations
with the opposite
signs. The anti--self-duality equations saturate the energy of {\it another} model,
namely that defined with the oppsite sign of $\kappa$ in \r{15}. This is
exactly what
happens in the case of the Higgs, or Ginzburg-Landau model, which clearly
explained in Ref. \cite{DI}.

Departing from this limiting case of saturated Bogomol'nyi bounds, we can
construct
models which would support non--self-dual vortices of arbitrary $N$,
provided that
the $O(3)$ breaking potential $U>U_0 =\frac{\mu^2}{2} (1-\phi^3)^2$ everywhere.
In that case we would
have to relax the temporal gauge and treat the component $A_0$ of $A_{\mu}$
in the
static field configuration as a dynamical coordinate together with
$(A_i,\phi^a)$ in
the Euler-Lagrange equations, as in \cite{BH}. We do not elaborate on such
details in the present Letter and suffice by noting the main features of
these vortex solutions.

The most remarkable feature of these vortices is that the lower
bound on their energy is given by a topological charge that is unrelated to
the magnetic flux but is in fact the usual degree of the map for the
corresponding ungauged $O(3)$ sigma model
\begin{equation}
\label{13}
\varrho_0
= \frac{1}{8\pi} \varepsilon_{ij} \varepsilon^{abc} \partial_i \phi^a
\partial_j \phi^b \phi^c
\end{equation}
This is shown in Ref. \cite{S} and demonstrated in details in Ref. \cite{ATY}. 
As a consequence, the value of the magnetic flux of these
vortices is not quantised, inspite of the solution being topologically stable.
The topological charge density \r{13} is not gauge invariant and hence
strictly speaking inadequate to supply a lower bound to the energy density
which is a gauge invariant quantity. As explained in Ref. \cite{ATY}, it is
possible to define the gauge invariant topological charge density by adding a
suitable total  divergence term with vanishing surface integral, which renders
it gauge  invariant. Since we will need this expression of the gauge invariant
topological charge density $\varrho$ in our dynamical considerations
later, we quote it here
\begin{equation}
\label{14}
\begin{array}{rl}
\varrho =&\varrho_0 +\frac{1}{4\pi} \varepsilon_{ij} \partial_i [(\phi^3
-1)A_j]\\
\\
&=\frac{1}{8\pi} \varepsilon_{ij} \varepsilon^{abc} D_i \phi^a D_j
\phi^b \phi^c +\frac{1}{4\pi} B(\phi^3 -1) .
\end{array}
\end{equation}
Topological charge densities like \r{14}, whose volume integrals yield the
degree of the map rather than the Chern-Pontryagin index associated with
the gauge group, occur \cite{T} in $SO(d)$ gauged $O(d+1)$ sigma models in $d$
dimensions, for all $d$.

The model \r{15} supports self-dual solutions in the static limit, but the
discussion of the dynamical properties of these vortices given below
applies equally well to non--self-dual solutions of models departing from
\r{15} by relaxing the restrictive choice of the $O(3)$
breaking potential $U_0 (\phi^3)$ in \r{15} given by \r{8b}, in the manner
specified above. That way we
can avail of the topological inequalities established in Refs. \cite{S,ATY},
which in these cases awould not be saturated. The solutions of the Euler-
Lagrange
equations will then be subject also to the Gauss Law constraint \r{6}, and
involve
an additional function parametrising the field $A_0$, in exactly the same
way as in the second item of Ref. \cite{BH}. As we will not perform any
numerical integrations in the present Letter, we do not pursue this
directon here.

In what we have done above, constructing a non-relativistic
$U(1)$ gauged $O(3)$ model for which we can state a topological
lower bound on the energy of the static field configuration, the most
important step is the determination of the function
$g(\phi^3)$. The particular choice of potential $U_0$ given by \r{8b} is
however not obligatory and serves only to enable the saturation of this bound.

\section{Dynamics}

Our final consideration is
the description of the dynamics of our vortices based on the formalism
developed in Ref. \cite{PT} for the vortices of the ungauged $O(3)$ sigma
model, and in Ref. \cite{PTST} for the analogous solitons of a non-
relativistic dynamical Ginzburg-Landau model. In both these examples,
the definition of the momentum field density depended crucially on the
gauge-invariant topological charge density that stabilises the vortex.
In the present example, this involves employing the topological charge
density defined by \r{14}.

Our purpose here is to verify that the prescriptions
introduced in Refs. \cite{PT,PTST}, namely that of employing the gauge-invariant
topological charge density in the definition of the momentum field density hold
also for the present model and are in that sense model independent.
It should be pointed out that  the conclusions of this section are
independent of
the form of the potential, and, the Chern-Simons term or the Maxwell term
might be
totally absent or be replaced by Lorentz non-invariant expressions
\footnote{In fact, since the rest of the Lagrangian is not Lorentz invariant
such terms will be induced by quantum effects.}.

For the presentation below,
it is convenient to use the angular representation $(\Theta, \Phi)$ of the
unit magnitude field $\phi^a$
\begin{equation}
\label{26}
\phi^1 =\sin\Theta \: \cos\Phi ,\qquad \phi^2 =\sin\Theta \: \sin\Phi ,
\qquad \phi^3 =\cos\Theta
\end{equation}
used in Ref. \cite{PT}.


Using formally the standard Noether
procedure for field configurations a) approaching their asymptotic values
fast enough for various surface terms arising in the variation of the action
to vanish and b) such
that all second spatial derivatives commute, one obtains the following
expression
for the field momentum density of the system:
\begin{equation}
\label{29}
p^N_i =\kappa \mu (1-\cos \Theta)(\partial_i \Phi -A_i) + \mu^2
\varepsilon_{ij} E_j B
\end{equation}
It is the sum of the momentum density
$\kappa \mu (1-\cos\Theta) (\partial_i \Phi - A_i)$ of the global model
properly covariantized to become gauge invariant and of the familiar
Poynting contribution $\varepsilon_{ij} E_j B$ of the pure Maxwell theory
\footnote{It is clear that the Chern-Simons term being metric-free does not
contribute to the energy-momentum tensor.}. Its volume integral
$P^N_i=\int d^2x p^N_i$ yields the momentum.

In the presence of a vortex the second spatial derivatives do not commute.
For a vortex with topological charge $N$ and located at ${\bf x}_0$ one
obtains $\epsilon_{ij} \partial_i \partial_j \Phi({\bf x}) = 2 \pi N
\delta({\bf x} - {\bf x}_0)$. As a consequence $P^N_i$ is not conserved.
$P^N_i$ is the correct expression for the momentum only in the absence of
topological solitons. We follow the steps described in Refs. \cite{PT} and
\cite{PTST} and write for the momentum of the model valid in all
topological sectors the formula:
\begin{equation}
\label{30}
P_i=\int d^2x \bigl[4 \pi \kappa \mu \varepsilon_{ij} x_j \varrho({\bf x}) +
\mu^2 \varepsilon_{ij} E_j B \bigr]
\end{equation}
where $\varrho$ is given by \r{14}.
It is conserved even in the presence of an arbitrary number of vortices
and antivortices.


The consequences of \r{30} for the dynamics of the
vortices are quite
surprising. Notice that the momentum of a static axially symmetric vortex
located at ${\bf a}$ and carrying topological charge $N$ is $4 \pi \kappa
\mu N \varepsilon_{ij} a_j$ and characterizes the position of the vortex
and not its motion. Correspondingly, for a vortex moving in formation with
velocity
${\bf v}$ one obtains $P_i=4 \pi \kappa \mu N \varepsilon_{ij} (a^o_j + v_j
t)$. Conservation of momentum implies $v_i=0$. Isolated vortices in the
absence of external forces are $\it spontaneously \; pinned$.

It is then natural to define the guiding centre ${\bf R}$ of an isolated
vortex of winding $N$
\begin{equation}
\label{33}
R_i=-{1\over 4 \pi \kappa \mu N} \varepsilon_{ij} P_j \end{equation}
Under a global displacement of the system by ${\bf c}$ it changes to ${\bf
R} + {\bf c}$ and for an axially symmetric vortex it coincides with its
geometric center. The above qualitative argument and the numerical study
of the equations of motion in similar systems \cite{PZST} leads to the
conclusion that ${\bf R}$ is a faithful representation of the mean position
of the vortex.
In the presence of an external force $F_j$ Newton's law $dP_j/dt = F_j$
translates into the vortex equation of mean motion \begin{equation}
\label{34}
{d\over dt} R_i = - {1\over 4 \pi \kappa \mu N} \varepsilon_{ij} F_j
\end{equation}
Generically, up to a fine cyclotron motion in its details \cite{PZST}, the
vortex moves as a whole with a speed proportional to the external force and
in a direction perpendicular to it.
It exhibits the so called Hall behaviour.

As a final important comment we would like to mention that it is possible to
develop the canonical
structure of the model and show that the momentum defined in \r{30} is
indeed the generator of spatial displacements satisfying the Poisson
brackets $\{P_k, {\cal F} \}= \partial_k {\cal F}$ with any field ${\cal F}$.
From this it follows that the two components of the momentum do not
commute. Instead, they satisfy
\begin{equation}
\label{35}
\{P_1, P_2 \} =4 \pi \kappa \mu N
\end{equation}
It is analogous to the Poisson bracket satisfied by the momentum of a
planar charged particle under the influence of a homogeneous
perpendicular magnetic field. Combined with the well-known fact that a
central extension of the linear momentum algebra is possible only for the
Euclidean algebra $E(2)$ in two dimensions and for the translational
algebra $T(D)$ in any dimension $D$, one concludes that the above Hall
motion of the solitons is only possible in two-dimensional systems with
$E(2)$ symmetry or in general $D-$ dimensional systems with just
translational invariance.

\section{Summary}

We have presented a non-relativistic $U(1)$ gauged $O(3)$ sigma model
with Maxwell and Chern-Simons dynamics which supports self-dual
winding number $N\ge 2$ vortices. This model reduces to the
Landau-Lifshitz theory of ferromagnetic materials in the ungauged limit.

It is possible to extend this model by making a different choice for the
$O(3)$ breaking potential, everywhere greater than the potential in \r{15},
resulting in non--self-dual vortices with arbitrary vorticity as in Ref.
\cite{GPS}. This generalisation is deferred to some future work.

We have shown that the dynamics of our vortices exhibits the Hall behaviour
familiar
in the  motion of planar charges in a perpendicular magnetic field
\cite{PTST}, as well
as in the motion of solitons in the global O(3) model and in the
non-relativistic
Maxwell gauged Ginsburg-Landau model \cite{PT}. In showing this we have
empoloyed the
prescription used in Refs. \cite{PT,PTST}, namely that of defining the
momentum field
density of the soliton in terms of the gauge invariant topological charge
density
\r{14}, in our definition of the momentum field density \r{30}.
We conclude therefore that this prescrition is model independent.

Our model exhibits nearly all the qualitative features of the
Ginzburg-Landau model,
except that in the limit of vanishing Maxwell term it is not possible to find
{\it zero energy} self-dual solitons as in the analogous non-linear
Schr\"odinger,
or non-relativistic Ginzburg-Landau, system \cite{JP}.

\end{document}